# THE LIMITED INTEGRATOR MODEL REGULATOR
# AND ITS USE IN VEHICLE STEERING CONTROL

Bilin Aksun-Guvenc, Levent Guvenc

## Abstract

Unexpected yaw disturbances like braking on unilaterally icy road, side wind forces and tire rupture are very difficult to handle by the driver of a road vehicle, due to his/her large panic reaction period ranging between 0.5 to 2 seconds. Automatic driver assist systems provide counteracting yaw moments during this driver panic reaction period to maintain the stability of the yaw dynamics of the vehicle. An active steering based driver assist system that uses the model regulator control architecture is introduced and used here for yaw dynamics stabilization in such situations. The model regulator which is a special form of a two degree of freedom control architecture is introduced and explained in detail in a tutorial fashion whereby its integral action capability, among others, is also shown. An auxiliary steering actuation system is assumed and a limited integrator version of the model regulator based steering controller is developed in order not to saturate the auxiliary steering actuator. This low frequency limited integrator implementation also allows the driver to take care of low frequency steering and disturbance rejection tasks. Linear simulation results are used to demonstrate the effectiveness of the proposed method.

# 1. Introduction

A two degree of freedom controller that provides very good disturbance rejection along with a high level of robustness to unmodeled dynamics is presented in this paper and is modified and applied to vehicle steering control with the aim of compensating for unexpected yaw rotations of the vehicle during the panic reaction period of the driver. This paper, therefore, has the two aims of first presenting this control scheme in a tutorial fashion and, secondly, showing that it can be used successfully as a vehicle steering controller after appropriate modifications. This control scheme will, therefore, be introduced first, followed by an introduction to the active vehicle steering task that is considered. Early work on this control architecture can be found in Ohnishi (1987) and Umeno and Hori (1991). Even though some authors have called this control architecture a disturbance observer, this name will not be used here as this control architecture, although somewhat similar in its disturbance cancellation objective, is not an implementation of a state estimator for disturbances as in Mita *et al.* (1998) and Liu and Peng (2000). Out of the other names like model regulator and disturbance estimating/canceling filter, the name *model regulator* will be used here. This is a logical choice since the aim of this control method is to force the input-output behaviour of the actual system to follow that of a chosen model in the presence of external disturbances and model uncertainty. The term two degree of freedom controller is also used frequently to describe the model regulator since it is a specific method of implementing/designing a two degree of freedom controller. The model regulator has been used successfully in a variety of motion control applications including high speed direct drive positioning (Kempf and Kobayashi, 1999) and friction compensation (Güvenç and Srinivasan, 1994). It can also be used to achieve dynamics similar to those obtained by the use of inner feedback loops through control (Güvenç, 1999). The model regulator has been applied to steering control in automated path following of a city bus (an IFAC benchmark problem) by Aksun Güvenç and Güvenç (2002). In automated path following, the steering controller has full control of the vehicle and guides the vehicle on a designated lane with magnetic markers. This is fundamentally different from the vehicle yaw stabilization task considered here where control authority is taken over only when needed, to make sure that the vehicle maintains yaw stability.

The aim in vehicle yaw stabilization through active steering control is to counteract undesired yaw rotations of the vehicle arising from unsymmetrical dynamic loading like braking on partially icy road, side wind or a tire blowout. In such situations, the driver cannot respond instantaneously. A dead time called the *panic reaction time* of the driver passes before the driver can start to take counter measures through steering or braking. This panic reaction time is about 0.5 to 2 seconds according to Ackermann (2000). So, the controller used in vehicle yaw stabilization should act before this 0.5 to 2 second period when a yaw moment disturbance occurs.

While one solution approach is to use individual wheel braking, the other solution approach of using a steering controller is preferred here[1]. The first approach, i.e. an individual wheel braking type vehicle yaw stabilization system, is available in production cars (see van Zanten, 2000, for example). The pioneering work in the second approach uses a robust decoupling based steering controller that works during the panic reaction period of the driver and counteracts the undesired yaw rotation (Ackermann, 1997). The same approach is taken here, except that a different steering controller architecture that makes use of the model regulator is used. The presence of an auxiliary steering system in which the auxiliary steering actuator adds small corrections to the steering angle commanded through the mechanical steering linkage by the driver is assumed. The aim is to let the driving task to the driver except during his panic reaction period following the occurrence of a yaw moment disturbance, during the course of which the active steering controller will impart its corrective action through the auxiliary steering actuator. This is in accordance with the work reported in Ackermann and Bünte (1997) and Aksun Güvenç *et al.* (2001a). One also needs to be careful in not saturating the auxiliary steering actuator and should therefore limit the controller action as in Ackermann and Bünte (1997). Therefore, the limited integrator implementation of the model regulator is introduced and used in this paper.

---

[1] Future driver assist systems for yaw stabilization are expected to use a combination of both approaches.

The organization of the paper is as follows. The conventional model regulator is presented and its structure is explained in detail in Section 2 which can also be viewed as a tutorial section on the model regulator. The single track car model used is introduced and numerical data being used along with the problem specifications are given in Section 3. The application of the model regulator to vehicle steering control for yaw stabilization is presented in Section 4. In this section, an auxiliary steering actuation system is assumed for active steering control and the limited integrator version of the model regulator based steering controller is developed. The paper ends with conclusions.

## 2. The Structure of the Model Regulator

The steering controller that is used here is based on the model regulator whose conventional architecture is shown in Figure 1 and comprises of the blocks $Q$ and $Q/G_n$ that form the two model regulator degrees of freedom. The augmentation of the plant *G* with the model regulator as seen in Figure 1 forces it to behave like its nominal (or desired) model within the bandwidth of the model regulator.

Using a verbal explanation, the model regulation in Figure 1 is achieved by first comparing the input u to the plant represented by $u_Q$ (approximately the same as u as will be seen later) with the input $u_y$ that should have been applied to obtain the measured output y based on the nominal (or desired) knowledge of the plant and then by passing the difference $u_{m.r.}$ through a positive feedback loop.

How this is done is explained more analytically, next. Consider plant *G* with multiplicative model error $\Delta_m$ and external disturbance d. Its input-output relation can be expressed as

$$y = Gu + d = (G_n(1+\Delta_m))u + d \tag{1}$$

where $G_n$ is the nominal (or desired) model of the plant. The aim in model regulator design is to obtain

$$y = G_n u_n \tag{2}$$

as the input – output relation in the presence of model uncertainty and external disturbance. $u_n$ in (2) is a new input signal to be defined below. This aim is achieved in model regulator design by treating the external disturbance and model uncertainty as an extended disturbance *e* and solving for it as

$$y = G_n u + (G_n \Delta_m u + d) = G_n u + e \tag{3}$$

$$e = y - G_n u \tag{4}$$

and using the new control signal $u_n$ given by

$$u = u_n - \frac{e}{G_n} = u_n - \frac{1}{G_n}y + u \tag{5}$$

to cancel the effect of *e* when substituted back in (3). With the aim of trying to limit the compensation to a pre-selected low frequency range (in an effort not to overcompensate at high frequencies and to avoid stability robustness problems), the feedback signals in (5) are multiplied by the low pass filter $Q$ which can also be viewed as a tunable design entity. In this case, the implementation equation becomes

$$u = u_n - \frac{Q}{G_n}(y+n) + Qu \tag{6}$$

where y+n with n representing the sensor noise is used as this is the actual output signal that is available. This was illustrated in the block diagram of Figure 1. The relative degree of the unity d.c. gain low pass filter $Q$ is chosen to be at least equal to the relative degree of $G_n$ for causality of $Q/G_n$. The loop gain of the model regulator compensated plant is

$$L = \frac{GQ}{G_n(1-Q)} \tag{7}$$

with the model regulation, disturbance rejection and sensor noise rejection transfer functions being given by

$$\begin{aligned}
\frac{y}{u_n} &= \frac{G_n G}{G_n(1-Q) + GQ} \\
\frac{y}{d} &= \frac{1}{1+L} = \frac{G_n(1-Q)}{G_n(1-Q) + GQ} \\
\frac{y}{n} &= \frac{-L}{1+L} = \frac{-GQ}{G_n(1-Q) + GQ}
\end{aligned} \tag{8}$$

from which it is obvious that $Q$ must be a unity gain low pass filter. As is desired, this choice will result in $y/u_n \rightarrow G_n$ (model regulation), $y/d \rightarrow 0$ (disturbance rejection) at low frequencies where $Q \rightarrow 1$, and $y/n \rightarrow 0$ (sensor noise rejection) at high frequencies where $Q \rightarrow 0$. Thus, conventional model regulator design is one of shaping *Q* as a low pass filter as illustrated in Figure 2. Of course, high frequency disturbance rejection is not possible with this choice, i.e. y/d→1 as Q→0 at high frequencies. This is not a big concern as such high frequency plant disturbances of significant size do not occur in vehicle yaw stabilization tasks. If one insists on using the model regulator for a general controls task incorporating high frequency plant disturbances, either the cutoff frequency of the Q filter can be increased along with better modeling and a higher fidelity sensor to achieve Q→1 at the high frequencies of interest or a disturbance feedforward compensation scheme can be utilized if it is possible to sense the high frequency disturbance.

A bandwidth limitation for the *Q* filter and hence the model regulator comes from the stability robustness requirement. To see this, rewrite the characteristic equation in (8) given by

$$G_n(1-Q) + G_n(1+\Delta_m)Q = 0 \tag{9}$$

as

$$G_n(1+\Delta_m Q) = 0 \;\rightarrow\; Q = -\frac{1}{\Delta_m} \tag{10}$$

An application of the small gain theorem (see Skogestad and Postlethwaite, 1996) results in

$$|Q| < \left|\frac{1}{\Delta_m}\right|, \;\; for \;\forall \omega \tag{11}$$

as the sufficient condition for robust stability. This requirement has also been illustrated in Figure 2. How to obtain the worst case plot (or the stability robustness boundary) for $|\Delta_{m,max}|$ such that $|\Delta_m| < |\Delta_{m,max}|$ for all frequencies will not be presented here as this paper does not focus on robustness analysis and robust controller design. Answers to these questions can be found for the general case in textbooks like Skogestad and Postlethwaite (1996) and Ackermann *et al.* (2002). The computation of $|\Delta_{m,max}|$ for a similar vehicle dynamics control problem can be found in Bünte *et al.* (2001).

For specific choices of $Q(s)$, the model regulator introduces integrators into the loop, resulting in reduced steady state error for reference and disturbance inputs. To see how the integrators are introduced, consider the equivalent form of the model regulator in Figure 3 and consider $Q(s)$ of the form

$$Q(s) = \frac{\alpha_m s^m + \cdots + \alpha_{k+1} s^{k+1} + a_k s^k + a_{k-1} s^{k-1} + \ldots + a_1 s + 1}{a_n s^n + a_{n-1} s^{n-1} + \ldots + a_2 s^2 + a_1 s + 1}, \quad k \leq m < n \quad (12)$$

Evaluate $Q/(1-Q)$ in the loop gain of (7) to obtain

$$\frac{Q}{1-Q} = \frac{\alpha_m s^m + \cdots + \alpha_{k+1} s^{k+1} + a_k s^k + a_{k-1} s^{k-1} + \ldots + a_1 s + 1}{s^{k+1} \left[ a_n s^{n-k-1} + a_{n-1} s^{n-k-2} + \ldots + (a_m - \alpha_m) s^{m-k-1} + (a_{m-1} - \alpha_{m-1}) s^{m-k-2} + \cdots + (a_{k+1} - \alpha_{k+1}) \right]} \quad (13)$$

which shows that k+1 integrators are incorporated into the loop for the choice of $Q$ in (12). Note that several commonly used forms for $Q(s)$ are of the form (12) and hence provide integral action. $Q(s) = 1/(\tau s + 1)^l$ (*m=0, n=l, k=0*) for example introduces one integrator into the loop. Similarly $Q(s) = 1/((s/\omega)^2 + (2\zeta/\omega)s + 1)$ (*m=0, n=2, k=0*) also introduces one integrator into the loop. $Q(s) = (3\tau s + 1)/(\tau^3 s^3 + 3\tau^2 s^2 + 3\tau s + 1)$ (*m=1, n=3, k=1*) used in Kempf and Kobayashi (1999) introduces two integrators into the loop. It should be noted that while the presence of integral action is useful for reducing steady state error, the model regulator is not simply an alternative design of an integral controller as suggested in Mita *et al.* (1998). Not all forms of the low pass filter $Q(s)$ will result in integral action. Certainly, the ideal situation of $Q=1$ does not result in integral action in the loop, but rather infinite gain at all frequencies.

## 3. Vehicle Model for Yaw Dynamics and Problem Specifications

The vehicle model which is used in this paper is the classical single track model, also called the bicycle model, which is shown in Figure 4. The main advantage of and reason for using the single track model is its simplicity. Its main disadvantages, of course, are that the four wheel distribution is not handled directly and that the roll and pitch motions are not modeled. However, keeping in mind its shortcomings, the single track model has been used extensively especially in the context of vehicle lateral control systems as is also done here. The major variables and geometric parameters of the single track model are

| | |
|---|---|
| $F_f$ $(F_r)$ | : Lateral wheel force at front (rear) wheel |
| $M_z$ | : Moment about the vertical axis $z$ through the center of gravity (CG) |
| $r$ | : Yaw rate |
| $\beta$ | : Chassis side slip angle at vehicle CG |
| $\alpha_f$ $(\alpha_r)$ | : Front (rear) tire side slip angle |
| $v$ | : Magnitude of velocity vector at CG ($v > 0, dv/dt = 0$) |
| $l_f$ $(l_r)$ | : Distance from front (rear) axle to CG |
| $\delta_f$ | : Front wheel steering angle |
| $m$ | : The mass of the vehicle |
| $J$ | : The moment of inertia w.r.t. a vertical axis through the CG |
| $c_f$ $(c_r)$ | : Front (rear) wheel cornering stiffness |
| $\mu$ | : friction coefficient at road – tire interface |

The transfer function from the front wheel steering angle $\delta_f$ to the yaw rate $r$ for the linearized single track model is given by (see Ackermann *et al.*, 2002)

$$G_{r\delta_f}(s) = \frac{r(s)}{\delta_f(s)} = \frac{b_1 s + b_0}{a_2 s^2 + a_1 s + a_0} \qquad (14)$$

with

$$b_0 = c_f c_r (l_f + l_r) v$$
$$b_1 = c_f l_f m v^2$$
$$a_0 = c_f c_r (l_f + l_r)^2 + (c_r l_r - c_f l_f) m v^2$$
$$a_1 = (c_f (J + l_f^2 m) + c_r (J + l_r^2 m)) v$$
$$a_2 = J m v^2$$

$G_{r\delta_f}(s)$ in (14) is also called the steering wheel input response transfer function here. The aim in active steering control is to shape this transfer function to achieve good steering command handling. The d.c. gain of the nominal single track model is

$$K_n(v) = \lim_{s \to 0} G_{r\delta_f}(s) \bigg|_{\mu = \mu_n} \qquad (15)$$

at the chosen longitudinal speed $v$ and at nominal friction coefficient $\mu = \mu_n$ (taken as unity here). The yaw disturbance input transfer function from yaw disturbance moment $M_z$ to yaw rate $r$ for the linearized single track model is given by

$$G_{rM_z}(s) = \frac{r(s)}{M_z(s)} = \frac{m v^2 s + (c_f + c_r) v}{a_2 s^2 + a_1 s + a_0} \qquad (16)$$

Note that a step input of $M_z$ can be used to model entering unilaterally icy road. The aim in active steering control is to reject the effect of yaw moment disturbances on the yaw motion of the vehicle (i.e. keeping corresponding yaw rate r small). The block diagram of the linearized single track model is shown in Figure 5. The reader is referred to references (see Ackermann *et al.*, 2002, for example) for more detailed information on the single track model. Note that the vehicle model given by equations (14) and (16) is a linear parameter (velocity v here) varying model.

The vehicle model data used here corresponds to a mid-sized passenger car that is available in the literature. The nominal values used are $l_f$=1.25 m, $l_r$=1.32 m, $m$=1296 kg, $J$=1750 kgm$^2$, $c_{f0}$=84000 N/rad and $c_{r0}$=96000 N/rad. The nominal friction coefficient is $\mu_n = 1$ which corresponds to dry road. The two variables exhibiting the largest variation during operation are the longitudinal speed $v$ and the mass $m$ of the vehicle. The tire cornering stiffnesses $c_f=\mu c_{f0}$ and $c_r=\mu c_{r0}$ can also exhibit large variations due to variations in friction coefficient $\mu$ between the road and the tires.

## 4. Model Regulator Based Steering Control

The model regulator based steering controller is illustrated in Figure 6. $G_{r\delta_f desired}$ is the desired steering command transfer function in this figure. $G_{sa}$ represents the auxiliary steering actuator. $G_{sa}$=1 is used here for easier presentation of results. The presence of an auxiliary steering actuator with a necessarily limited range of action is taken into account indirectly in the model regulator based steering controller design reported here by using limited integrator action. In previous work, $Q$ and $G_{r\delta_f desired}$ were chosen as

$$Q = \frac{1}{\tau_Q s + 1} \tag{17}$$

$$G_{r\delta_f desired} = \frac{K_n(v)}{\tau_n s + 1} \tag{18}$$

(see Bünte *et al.*, 2001, and Aksun Güvenç *et al.*, 2001b, respectively) to achieve integral action for the steer-by-wire type implementation of the model regulator based steering controller. (18) requires the use of a velocity scheduled implementation of the model regulator which was used successfully in Aksun Güvenç *et al.* (2001a). To avoid possible steering actuator saturation problems that are much more demanding when auxiliary steering actuation is considered, a band pass *Q* filter was used in Aksun Güvenç *et al.* (2001a). The approach taken here is more direct and uses a limited integrator type implementation of the model regulator in order not to overload the steering actuator. This is similar to the fading integrator idea of Ackermann and Bünte (1997).

Regarding the choice of a first order form for the desired dynamics $G_{r\delta_f desired}$ in (18) as compared to second order model dynamics $G_{r\delta_f}$ in (14), there are three reasons for this choice. They are: 1) to avoid an oscillatory controlled yaw response, 2) to show that a lower order dynamic behavior can be imposed on the system by using the model regulation concept and 3) to have a simple controller at the end. It is, of course, possible to use a second order form similar to (14) but with nominal parameter values and with $\mu=\mu_n=1$ (dry road) instead of (18) for $G_{r\delta_f desired}$. It is also possible to use a critically damped second order transfer function for $G_{r\delta_f desired}$.

## 4.1. Limited Integrator Implementation

Another practical problem that needs to be addressed is one of man-machine interface. The task of the steering controller is to intervene only during the panic reaction period of the driver when a yaw moment disturbance occurs and to hand over the steering task back to the driver afterwards. In control specific terms, the steering controller should only react to high frequency inputs outside the bandwidth of the driver. This is called short duration steering control here. Note that this effect can also be called the fading effect and is necessary when an auxiliary steering controller is used in order to avoid saturating the auxiliary steering actuator with a limited capability of steering wheel angle change.

Recall from Section 2 that the choice (17) for $Q$ results in an integrator being incorporated into the loop gain as

$$\frac{Q}{1-Q} = \frac{1}{s}R(s) = \frac{n_R(s)}{d_R(s)}. \tag{19}$$

where R(s) represents everything other than the integrator. This results in zero steady state error in response to step reference and disturbance inputs. Short duration steering control can be achieved by using a low frequency limited integrator as illustrated in Figure 7 instead of an integrator in the loop. For this purpose, it is necessary to incorporate the low frequency limited integrator $K/(\tau s + 1)$ instead of $1/s$ in the model regulator loop using

$$\frac{Q}{1-Q} = \frac{K}{\tau s+1}R(s) = \frac{K}{\tau s+1}\frac{n_R(s)}{d_R(s)}. \tag{20}$$

Back solving for $Q$ in (20) results in

$$Q = \frac{Kn_R(s)}{(\tau s+1)d_R(s) + Kn_R(s)} \tag{21}$$

Using the simplest possibility with $R=n_R=d_R=1$ results in

$$Q = \frac{K}{\tau s+1+K} = \frac{K/(1+K)}{\left(\frac{\tau}{1+K}\right)s+1} \tag{22}$$

The simulation results with (22) as $Q$, with $K=10$, $\tau=0.006$ sec and with $G_{r\delta_f desired}$ given in (18) are shown in Figures 8 and 9. The simulations were conducted at six operating conditions corresponding to different values of speed and road–tire friction coefficient, the latter ranging from icy to dry road. Step commands of driver steering input (see Figure 8) and yaw moment disturbance input (see Figure 9) were treated in these simulations. The responses with the active steering controller of this section are displayed using solid lines (controlled) and dashed lines are used in its absence (conventional) in Figures 8 and 9. It is seen from an examination of these figures that the short duration steering intervention objective has been met. The steering command responses in Figure 8 follow the desired steering dynamics specified in (18) quite well at all operating conditions. In interpreting the results displayed in Figure 8, the following remarks should be taken into account: 1) the steering wheel step input value has been normalized to obtain unity as the steady state value of the yaw rate under dry road condition (i.e. μ=1), 2) the conventional vehicle yaw rate responses on the right hand side of Figure 8 exhibit large values of deviation from unity at steady state as the road is not dry (μ<1), 3) the controlled vehicle yaw rate responses on the right hand side of Figure 8 have values close to unity at steady state, the small deviations from unity being due to the limited integrator nature of the model regulator used.

The disturbance rejection is also very good with the model regulator with low frequency limited integrator action as is seen in Figure 9 but a steady state error, which can easily be zeroed by the driver, remains. An advantage of the model regulator based steering controller is that it is easily adjusted for satisfying different goals. The desired steering dynamics can be specified for better handling dynamics, for instance, by changing $G_{r\delta_f desired}$ in (18). The disturbance rejection properties and the amount of limited integrator action can be adjusted by modifying $K$ and $\tau$ in (22). The design can also be carried out in the space of chosen controller parameters to satisfy frequency response magnitude constraints like weighted sensitivity or mixed sensitivity bounds (see Bünte *et al.*, 2001, for example).

Note that only constant velocity operation at three different velocities has been treated here for ease of exposition. In a practical implementation, the controller is also velocity scheduled ($K_n(v)$ changes with velocity in equation (18)). This has been illustrated with a different approach to Q filter design in Aksun Güvenç *et al.* (2001a).

Note that one of the aims in using limited integrator type model regulation was to try to reduce the risk of auxiliary steering actuator saturation. The maximum value of the yaw disturbance moment is taken as 4000 Nm in this study and has been used in the simulations of Figure 9. The auxiliary steering actuator saturation limit is specified as corresponding to ±3° of wheel rotation here. The minimum value of the disturbance moment that causes the auxiliary steering actuator to saturate turns out to be 6119 Nm. The auxiliary steering actuator did not saturate during any of the simulations reported here. For purposes of illustration, the auxiliary steering actuator output $\delta_{m.r.}$ is shown in Figure 10 for both the limited integrator model regulator (solid plot) discussed until now and a standard model regulator (dashed plot). For the standard model regulator, the only difference was the use of Q(s) in (17) with $\tau_Q=\tau/(1+K)$ instead of equation (22). Comparison of the two cases in Figure 10 shows that the standard model regulator is closer to the limit of saturation ($|\delta_{m.r.}|_{max}$=2.24°) in comparison to the limited integrator one ($|\delta_{m.r.}|_{max}$=1.86°). While a properly configured Q(s) will reduce the risk of saturation, there will always be a large enough disturbance that will cause saturation. The consequence of auxiliary steering actuator saturation will be a degradation in performance. Therefore, one should be careful to avoid the possibility of saturation in design. This can be achieved by using a less aggressive controller or by using an auxiliary steering actuator with a larger saturation limit.

## 5. Conclusions

Active vehicle steering controller development for yaw dynamics stabilization was treated in this paper. An auxiliary steering actuation system was assumed and a low frequency limited integrator version of a model regulator based steering controller was developed in order not to saturate the auxiliary steering actuator and to let the driver take care of low frequency steering and disturbance rejection tasks. Linear simulation results have shown the effectiveness of this method for improving vehicle steering dynamics and in rejecting yaw moment disturbances. Further work should use a realistic steering actuator model and a higher order, more realistic vehicle model for simulation.

More recent and other work of the authors on disturbance observer (or model regulator) control, yaw stability control applications and extension to automated vehicles can be found in Aksun-Guvenc et al (2003, 2009), Oncu et al (2007), Ding

et al (2020), Emirler et al (2018), Ma et al (2020, 2021), Guvenc et al (2021, 2017), Zhu et al (2020, 2019), Wang et al (2018), Gelbal et al (2020), Hacibekir et al (2006).

**Acknowledgment**

The second author acknowledges the support of the Alexander von Humboldt Foundation.

Wang, H., Tota, A., Aksun-Guvenc, B., Guvenc, L., 2018, "Real Time Implementation of Socially Acceptable Collision Avoidance of a Low Speed Autonomous Shuttle Using the Elastic Band Method," IFAC Mechatronics Journal, Volume 50, April 2018, pp. 341-355.

van Zanten, A.T., "Bosch ESP Systems: 5 Years of Experience," SAE paper no. 2000-01-1633, 2000.

Zhu, S., Aksun-Guvenc, B. Trajectory Planning of Autonomous Vehicles Based on Parameterized Control Optimization in Dynamic on-Road Environments. J Intell Robot Syst 100, 1055–1067 (2020). https://doi.org/10.1007/s10846-020-01215-y

Zhu, S., Gelbal, S.Y., Aksun-Guvenc, B., Guvenc, L., 2019, "Parameter-space Based Robust Gain-scheduling Design of Automated Vehicle Lateral Control," IEEE Transactions on Vehicular Technology, doi: 10.1109/TVT.2019.2937562, Vol. 68, Issue 10, pp. 9660-9671.22

**FIGURES**

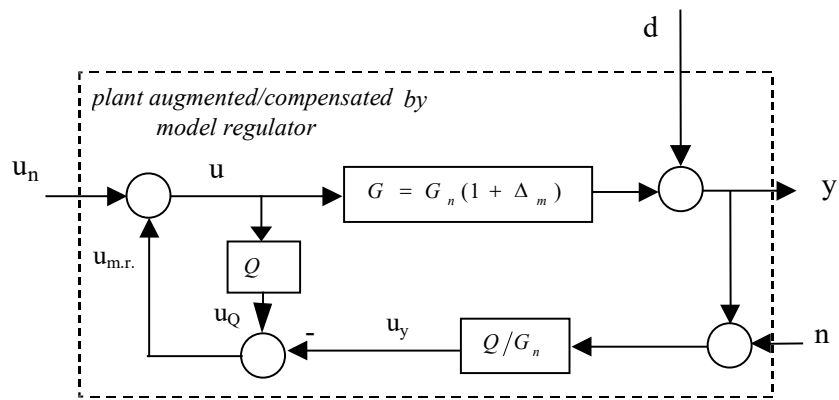

**Figure 1.** System with model regulator

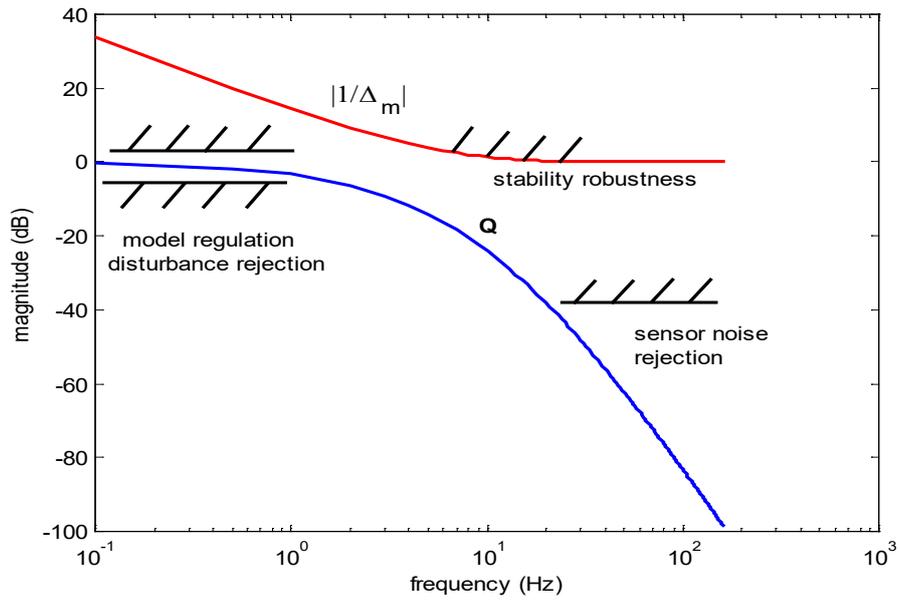

**Figure 2.** Model regulator design requirements

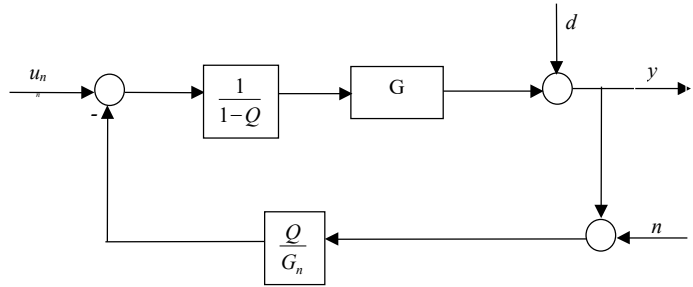

**Figure 3.** Equivalent form of model regulator

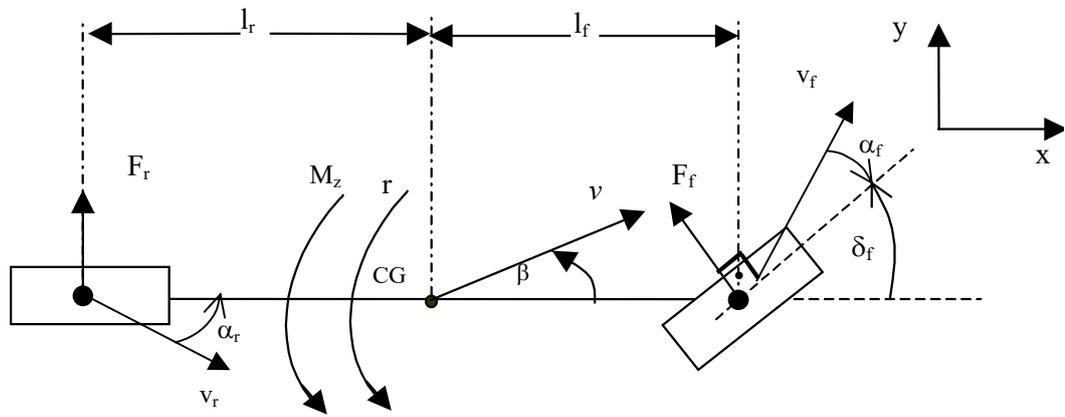

**Figure 4.** Single track model for car steering

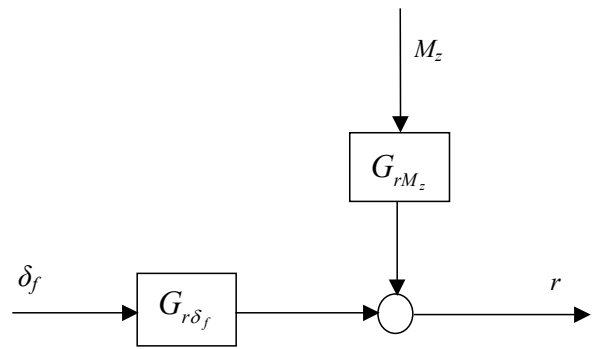

**Figure 5.** Linearized single track model

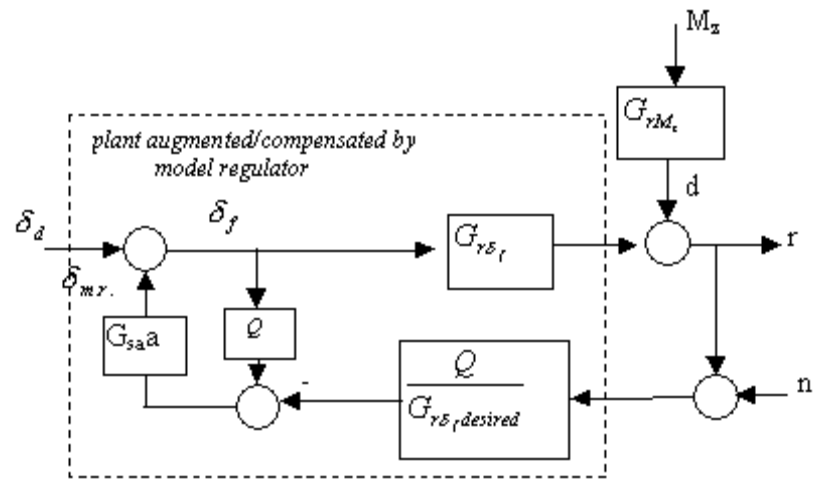

**Figure 6:** Model regulator based steering controller

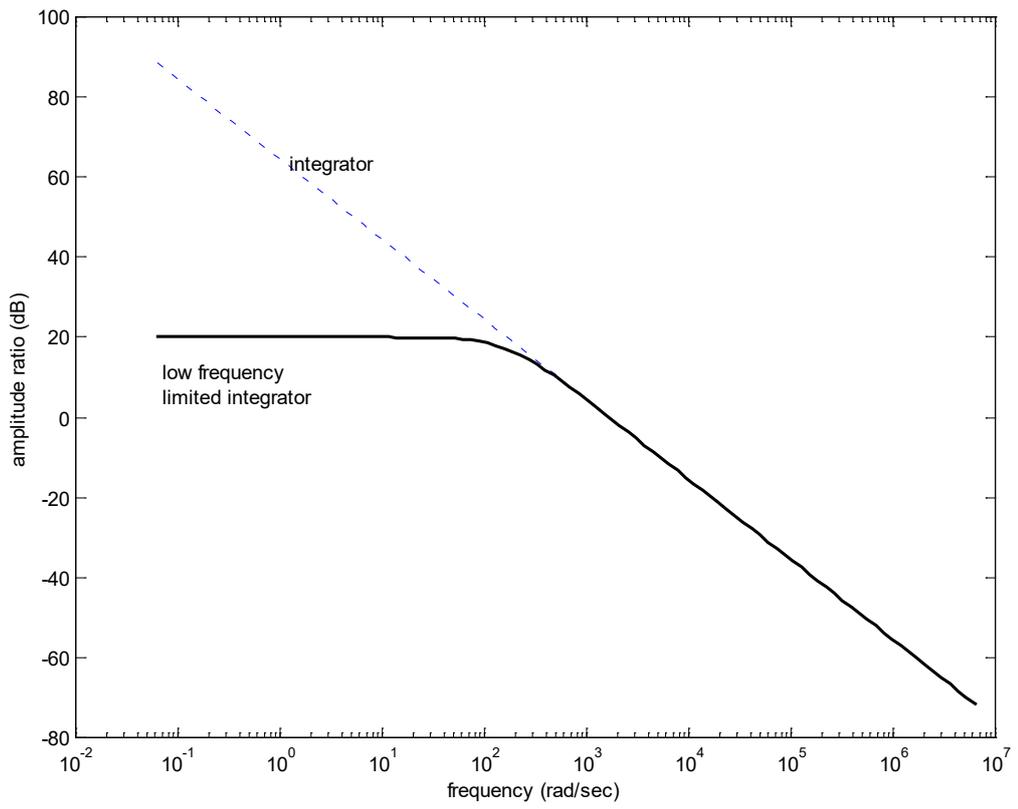

**Figure 7.** Low frequency limited integrator

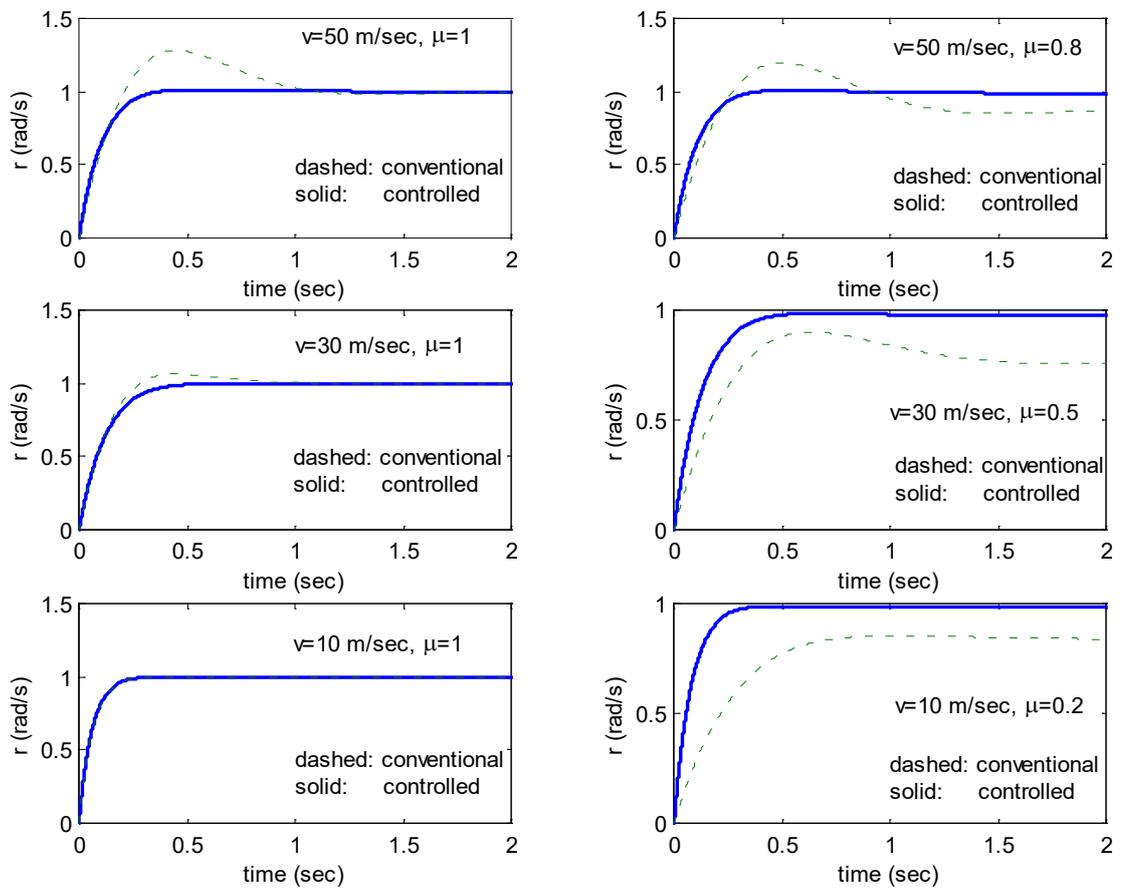

**Figure 8.** Steering wheel step input responses

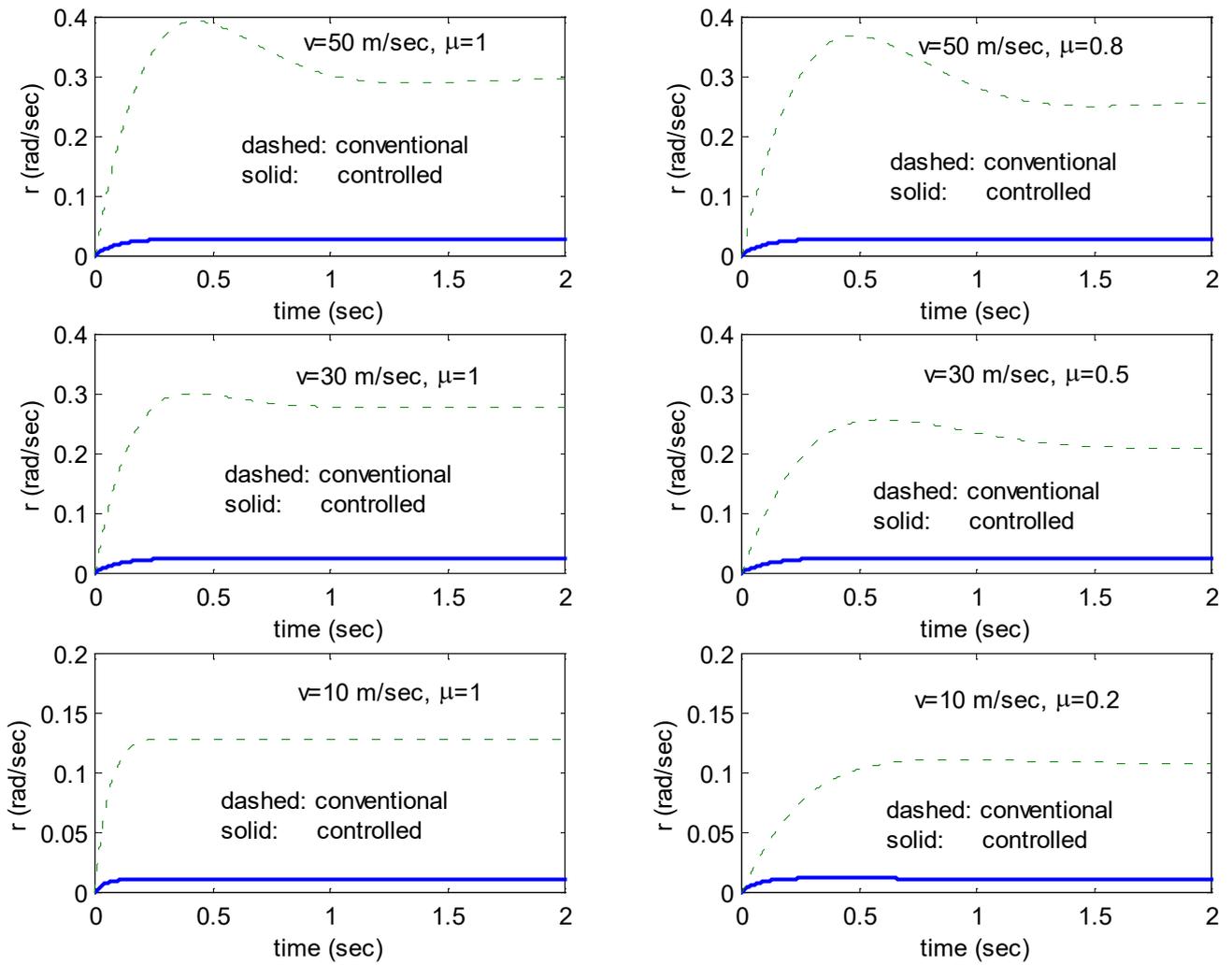

**Figure 9.** Yaw moment step disturbance responses

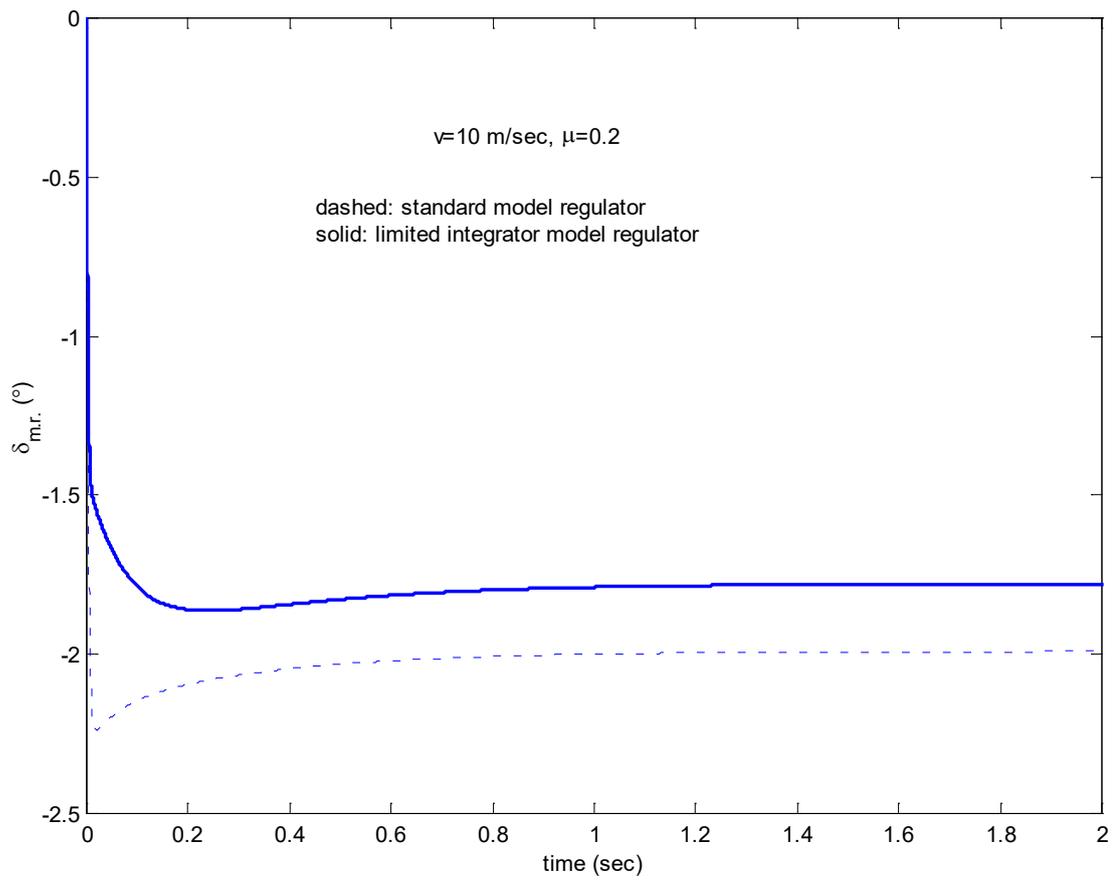

**Figure 10.** Auxiliary steering actuator output comparison